\documentclass[twocolumn,letterpaper,nofootinbib,prd,amsmath]{revtex4}

\usepackage[dvips]{graphics,color}
\usepackage{epsfig}
\usepackage{hyperref}
\newcommand{\RR}{\hbox{$I$\kern-3.8pt $R$}}

\begin{document}

\title{Distorted Black Hole Initial Data
    Using the Puncture Method}  
\author{J.~David Brown}
\affiliation{Department of Physics, North Carolina State University,
Raleigh, NC 27695 USA}
\author{Lisa L.~Lowe}
\affiliation{Department of Physics, North Carolina State University,
Raleigh, NC 27695 USA}

\begin{abstract}
We solve for single distorted black hole initial data using the puncture method, where the 
Hamiltonian constraint is written as an elliptic equation in $\RR^3$ for the nonsingular 
part of the metric conformal factor. With this approach we can generate isometric and non--isometric 
black hole data. For the isometric case, our data are directly comparable to those
obtained by Bernstein {\it et al.}, who impose isometry boundary conditions 
at the black hole throat.  Our numerical simulations are performed using a parallel multigrid 
elliptic equation solver with adaptive mesh refinement. Mesh refinement allows us to 
use high resolution around the black hole while keeping the grid boundaries far away in the 
asymptotic region. 
\end{abstract}
\maketitle

\section{Introduction}
Distorted black holes are expected to be produced by astrophysical events such as asymmetrical gravitational 
collapse, black hole--black hole coalescence, black hole--neutron star coalescence, and possibly neutron 
star--neutron star coalescence. As ground based gravitational wave detectors such as LIGO, VIRGO, 
TAMA, and GEO increase their sensitivities, and the planned launch date for the space based detector 
LISA grows near, the need for researchers to develop a theoretical understanding of these systems 
becomes increasingly important. In support of that effort, we need to develop the techniques and tools to evolve 
a dynamic, distorted black hole and extract the emitted gravitational radiation as it settles to a quiescent 
state.

One of the difficulties encountered in the numerical treatment of single and multiple black hole systems 
is the presence of nontrivial topologies. In the past, researchers have addressed this problem by 
using isometry conditions at inner boundaries to represent black hole throats. Another 
approach to black hole evolution is excision, in which a region inside 
each apparent horizon is removed from the computational domain. Recently it has been 
shown that black holes can be treated in terms of fields on $\RR^3$ by splitting the conformal 
factor for the spatial metric into singular and non--singular terms. This so--called ``puncture method'' 
was applied to the Bowen--York \cite{Bowen:1980yu} family of black hole initial data sets by 
Brandt and Br{\"u}gmann \cite{Brandt:1997tf}, and used for evolution studies by 
Br{\"u}gmann \cite{Brugmann:1997uc} and others (see, for example, Refs.~\cite{Alcubierre:2002kk,Imbiriba:2004}). 

In this paper we make a modest extension of the puncture construction for initial data 
to include single distorted black holes. We reproduce and extend the results of Bernstein 
{\it et al.}~\cite{Bernstein:1994ny,Brandt:1996si,Brandt:2002wa}, who constructed 
``black hole plus Brill wave'' initial 
data sets using isometry conditions at the black hole throat. Whereas the black holes obtained 
by Bernstein {\it et al.}~are isometric by construction, our distorted puncture black hole data sets
are isometric only when the free parameters $\mu$ and $m$, defined in Sec.{~\ref{sec2}}, are equal to one another. 
For $\mu \neq m$, we obtain non--isometric, distorted black holes. 

Another difficulty encountered in the numerical modeling of black holes is the wide discrepancy in 
length scales involved.  The computational grid needs to be large enough to capture outgoing gravitational 
waves and to minimize boundary effects.  The grid must also have high resolution in the 
interior to accurately resolve the strong gravitational fields of black holes. For a finite difference 
code, adaptive mesh refinement (AMR) is needed to satisfy both of these requirements:
high resolution and a large grid.  In this paper we introduce our AMR elliptic
solver that allows us to solve accurately for puncture black hole data on very large grids.  
Evolution studies of these data sets are underway \cite{Fiske}. 
 
In Sec.{~\ref{sec2}} we set up the equations defining the initial value problem for a distorted
puncture black hole.  We show how the puncture data can be formulated to give isometric data sets, thus enabling a
comparison  with the results of Bernstein {\it et al.} In Sec.~{\ref{sec3}} we give a brief 
description of our AMR elliptic solver. In Sec.~\ref{sec4} we present sample results both for 
isometric and non--isometric black hole data. 

\section{Formulation of the Problem}\label{sec2}
Following Bernstein {\it et al.}~\cite{Bernstein:1994ny,Brandt:1996si,Brandt:2002wa} 
we write the line element for the 
physical metric $g_{ij}$ as
\begin{equation} \label{eqn:brillmetric}
  ds^2 = \psi^4 [ e^{2q}(dr^2 + r^2\,d\theta^2) + r^2\sin^2\theta\, d\phi^2] \ ,
\end{equation}
where $q$ is a specified function of the spatial coordinates and $\psi$ is the conformal factor. 
In this paper we restrict ourselves 
to the expression for $q$ used in Ref.~\cite{Brandt:2002wa}, namely, 
\begin{equation} \label{eqn:qfunction}
  q(r,\theta,\phi) = 2 Q_0 e^{-\eta^2} \sin^n\theta\, (1 + c \cos^2\phi) \ ,
\end{equation}
where $\eta = \ln(2r/\mu)$. In Eq.~(\ref{eqn:qfunction}), $Q_0$, $n$, $c$ and $\mu$ are adjustable constants 
that affect the size and type of distortion. We also restrict ourselves to data sets 
defined at a moment of time symmetry. Thus, the extrinsic curvature vanishes and the 
momentum constraints are trivially satisfied. The Hamiltonian constraint reduces to 
\begin{equation} \label{eqn:HC}
  {\hat\nabla}^2 \psi - \frac{1}{8} {\hat R} \psi = 0 \ ,
\end{equation}
where $\hat\nabla^2$ and $\hat R$ are the Laplacian and scalar curvature of the conformal metric, 
defined by $\hat g_{ij} = \psi^{-4} g_{ij}$. 

Solutions of Eq.~(\ref{eqn:HC}) on the manifold $\RR^3$ are Brill waves. 
The distorted black hole data sets of Bernstein  {\it et al.}~were obtained by solving 
Eq.~(\ref{eqn:HC}) on the manifold $\RR\times S^2$, with Robin 
conditions at the outer boundary [$\partial(r\psi - r)/\partial r = 0$ as $r \to \infty$] 
and isometry conditions at the inner boundary
[$\partial(\sqrt{r}\psi)/\partial r = 0$ at $r = \mu/2$]. We will obtain solutions of 
Eq.~(\ref{eqn:HC}) on $\RR\times S^2$  
by following the puncture construction of Ref.~\cite{Brandt:1997tf}. Thus, we write the  conformal factor as
$\psi = u +  m/(2r)$ 
and insert this into the Hamiltonian constraint (\ref{eqn:HC}) to obtain  
\begin{equation} \label{eqn:ueqn}
  {\hat\nabla}^2 u - \frac{1}{8} {\hat R} u = \frac{m}{16 r} {\hat R} \ .
\end{equation}
This equation is solved for a continuous solution $u$ on $\RR^3$ with Robin boundary conditions 
$\partial(r u - r)/\partial r = 0$ at $r \to \infty$ . Note that the ``bare mass'' $m$ appears 
as a new parameter in the construction.\footnote{The parameters $\mu$ and $m$ are 
dimensionful. Thus, our data sets can be described in terms of the dimensionless 
parameters $c$, $Q_0$, $n$, the dimensionless ratio $\mu/m$, and the dimensionless coordinates $x/m$, 
$y/m$, and $z/m$.} 

We have had no 
difficulty in obtaining numerical solutions to Eq.~(\ref{eqn:ueqn}). 
However, from an analytical point of view, it is not immediately obvious to us whether or not Eq.~(\ref{eqn:ueqn}) 
always admits solutions, and if so whether those solutions are unique. 
The problem of existence and uniqueness of solutions of Eq.~(\ref{eqn:ueqn}) on $\RR^3$ is 
similar to the problem of existence and uniqueness of solutions of
Eq.~(\ref{eqn:HC}) on $\RR^3$. There are two notable differences. First is the 
appearance of a ``source'' term $m{\hat R}/(16 r)$ on the right--hand side of Eq.~(\ref{eqn:ueqn}). Note 
that with the choice (\ref{eqn:qfunction}) for the function $q$, the scalar curvature $\hat R$ 
goes to zero rapidly, $\hat R \sim (\ln(r))^2 r^{\ln(\mu/r) - 2 - 2\ln(2)} $ at $r \to 0$, 
so that $\hat R/r$ does not blow up at the puncture. The second difference 
between equation  (\ref{eqn:ueqn}) and the familiar initial value equation (\ref{eqn:HC}) is 
that, for (\ref{eqn:ueqn}), we 
need not require the solutions  to be positive. 
Rather, we would like to know if $u$ is greater than $-m/(2r)$ so that the combination 
$\psi = u +  m/(2r)$ is everywhere positive.

By construction the data sets of Bernstein {\it et al.}~are  isometric; that is, they are 
symmetric under reflections about the black hole throat $r = \mu/2$. More  precisely, 
each data set is described by a metric tensor whose 
components $g_{ij}$, as functions of $r$, $\theta$ and $\phi$, are unchanged 
by the coordinate transformation $r \to \bar r \equiv \mu^2/(4r)$.
We can solve for reflection symmetric 
data sets using the puncture method as well, simply by setting the parameters $\mu$ and $m$ equal to 
one another: $\mu = m$. To see this, we first note that the Hamiltonian constraint 
on $\RR\times S^2$ can be written as (see, for example, Appendix D of Ref.~\cite{Wald:1984rg}) 
\begin{equation} \label{eqn:HC2}
  \left( \tilde\nabla^2 - \frac{1}{8}\tilde R \right) (\sqrt{r} \psi) = 0 \ ,
\end{equation}
where $\tilde\nabla^2$ and $\tilde R$ are the Laplacian and scalar curvature for the 
metric $\tilde g_{ij} = {\hat g}_{ij}/r^2$. Let $\psi_1(r)$ denote a solution of 
Eq.~(\ref{eqn:HC}), or equivalently Eq.~(\ref{eqn:HC2}), 
obtained by the puncture method. (For notational simplicity, we 
display only the $r$ dependence in the solution $\psi_1$, and in $\psi_2$ below. In general 
these are functions of $\theta$ and $\phi$ as well.) This solution $\psi_1(r)$ has boundary 
behavior $\psi_1(r) \to 1$ 
as $r \to \infty$, and $\psi_1(r) \to m/(2r)$ as $r \to 0$. Now observe 
that the line element $ds^2 = \tilde g_{ij} dx^i dx^j$, with the function $q$ chosen as 
in Eq.~(\ref{eqn:qfunction}), is invariant under reflections 
$r \to \bar r = \mu^2/(4r)$. The scalar operator  $\tilde\nabla^2  - \tilde R/8$ is invariant as well.
By making the substitution 
$r \to \bar r = \mu^2/(4r)$ in Eq.~(\ref{eqn:HC2}) we see that $\psi_2(r)$, defined by 
$\sqrt{r}\psi_2(r) = \sqrt{\bar r} \psi_1(\bar r)$ with $\bar r = \mu^2/(4r)$, is also a solution of 
Eqs.~(\ref{eqn:HC2}) and (\ref{eqn:HC}). This solution satisfies the 
boundary conditions $\psi_2(r) \to m/\mu$ 
as $r \to\infty$ and $\psi_2(r) \to \mu/(2r)$ as $r \to 0$. If we choose $\mu = m$, then the solutions $\psi_1(r)$ 
and $\psi_2(r)$ satisfy the same equation and obey the same boundary conditions. If we assume that solutions 
to the puncture equation (\ref{eqn:ueqn}) are unique, then the two solutions $\psi_1(r)$ and $\psi_2(r)$ must in 
fact be identical. From the equality $\psi_1(r) = \psi_2(r)$ we find 
\begin{equation}\label{eqn:psirelation}
  \sqrt{r} \psi_1(r) = \left.\left( \sqrt{\bar r} \psi_1(\bar r) \right) \right|_{\bar r = \mu^2/(4r)} \ .
\end{equation}
The line element (\ref{eqn:brillmetric}) for this solution can be written as 
\begin{equation}\label{eqn:brillmetric2}
  ds^2 = (\sqrt{r} \psi_1(r))^4 [e^{2q}(dr^2/r^2 + d\theta^2) + \sin^2\theta \,d\phi^2] \ .
\end{equation}
Then the relation (\ref{eqn:psirelation}) is the condition for the physical metric (\ref{eqn:brillmetric2}) 
to be invariant under the reflection $r \to \bar r =  \mu/(4r)$.

In Sec.~{\ref{sec4} we present results for the conformal factor $\psi$ for isometric 
($\mu = m$) and non--isometric ($\mu \ne m$) data sets. For isometric data, we can check the reflection 
symmetry by comparing the ADM mass computed at the two infinities. At the ``outer'' infinity ($r \to\infty$) the 
ADM mass is given by \cite{OMurchadha:1974} 
\begin{equation}
  M_\infty = -\frac{1}{2\pi} \oint_\infty d{\hat S}^i\,\hat\nabla_i \psi \ .
\end{equation}
Numerically, the integral is computed at the outer boundary of the computational domain. 
The ADM mass at the ``inner'' infinity ($r=0$) is found by expressing the metric (\ref{eqn:brillmetric})
in coordinates $\bar r$, $\theta$, $\phi$, where $\bar r = m^2/(4r)$. We find
\begin{equation}
  ds^2 = \left( 1 + \frac{m u}{2\bar r} \right)^4 
    [ e^{2 q}(d\bar r^2 + \bar r^2\,d\theta^2) + \bar r^2\sin^2\theta\, d\phi^2] \ ,
\end{equation}
where $q$ and $u$ are functions of $r = m^2/(4\bar r)$, $\theta$, and $\phi$. Provided 
$q$ goes to zero sufficiently rapidly as $\bar r \to \infty$, the metric is asymptotically 
flat at $r = 0$ with ADM mass given by 
\begin{equation}
  M_0 = m \left. u \right|_{r = 0} \ .
\end{equation}
The choice (\ref{eqn:qfunction}) for $q$ vanishes as $r\to 0$ like $q \sim r^{\ln(\mu/r)}$, 
faster than any power of $r$.
\section{Numerical Code}\label{sec3}
Our numerical code solves Eq.~(\ref{eqn:ueqn}) using multigrid techniques with mesh refinement. 
We use the Paramesh package \cite{MacNeice00,parameshmanual} to implement 
mesh refinement and parallelization. Paramesh divides the computational grid into blocks, 
with each block containing
$N$ zones. A block of data is refined by bisection---that is, the block is divided into eight 
``child'' blocks (in three spatial dimensions) each containing $N$ zones. 

Our code carries out multigrid V--cycles using the Full Approximation Storage (FAS) algorithm 
\cite{NumericalRecipes} on non--uniform grid structures, 
with zone--centered data. It works with 
both fixed mesh refinement (FMR) and adaptive mesh refinement (AMR). Working with FMR, we specify a non--uniform 
grid by hand. 
Working with AMR, 
we generally start with a 
coarse, uniform grid. The code V--cycles until the norm of the residual is less than the norm of the  
relative truncation error in each block. Any block whose 
relative truncation error is above a specified tolerance is flagged for refinement. Paramesh then rebuilds 
the grid, the data is prolonged from the old grid to the new, and the V--cycle process begins again. Working in 
this mode takes the place of the Full Multigrid (nested iteration) Algorithm \cite{NumericalRecipes}, 
where the solution 
on the final grid structure is reached by a succession of V--cycles of increasing peak resolution. 

We will provide details of the AMR--multigrid algorithm in a later publication, along with 
numerous code tests \cite{Brown:2004ma}. 
\section{Results}\label{sec4}
Bernstein {\it et al.}~\cite{Bernstein:1994ny,Brandt:1996si,Brandt:2002wa}  
produced nonaxisymmetric distorted black holes using isometry conditions at the black hole throat.  
We have reproduced several of the initial data sets from Ref.~\cite{Brandt:2002wa} using the puncture 
method, by setting $\mu = m$ as described in Sec.~\ref{sec2}. As a check for isometry, we 
have confirmed that the ADM mass at the puncture, $M_0$, agrees with the ADM mass at infinity, 
$M_\infty$, to several significant digits.\footnote{A direct numerical comparison of 
our ADM masses with those of Bernstein {\it et al.} is not possible, since their ADM masses 
are presented graphically. A visual comparison our ADM mass data with theirs shows good agreement.} 

Two tests were performed to confirm the consistency of our ADM mass calculations. For the first test 
we used a sequence of grids with fixed interior resolution but increasing outer boundary limits.  
We started by solving for the initial data on a grid with boundaries (in each dimension) at $-13$ and $13$, 
and having two additional 
box--in--box refinement levels ranging (in each dimension) from $-6.5$ to $6.5$ and from $-3.25$ to $3.25$. 
The resolution on the finest of the three levels was $\Delta x = 0.05078$.  Other grids in the sequence
were created by adding, one at a time, another fixed box--in--box refinement level while doubling the 
size of the computational 
domain. This test was intended primarily as a check on the sensitivity of the ADM mass at infinity to the location of 
the outer boundary. We found that the ADM mass $M_\infty$ is unaffected to six digits and the ADM mass $M_0$ is unaffected to seven digits. 
For the second test we used a sequence of grids consisting of a 
fixed three--level box--in--box 
structure but increasing resolution throughout the computational domain. This test was intendend primarily 
to check the convergence properties of the ADM masses $M_\infty$ and $M_0$. The test showed that the 
ADM masses converge with second order accuracy. 

As a specific example let us consider the data set discussed in Ref.~\cite{Brandt:2002wa} 
with $c=-2$, $Q_0 = -0.5$, $n = 4$, and $\mu = m = 2$. The ADM masses for this data at 
infinity and at the puncture are $M_\infty = 2.2021$ and $M_0 = 2.2027$.   
Figure \ref{C=-2.Q=-0.5.-100to100.tau=0.005.2inone} shows  a contour plot of the nonsingular 
part $u$ of the conformal factor in the $z = 0$ plane.
\begin{figure}
\includegraphics{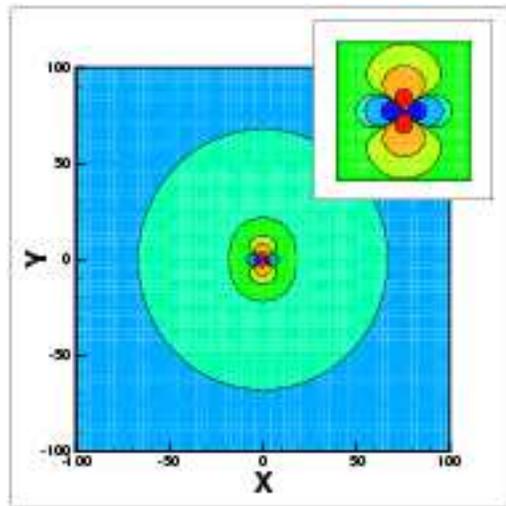}
\caption{Contour plot of $u$ in the $z=0$ plane for the case $c=-2$, $Q_0 = -0.5$, $n = 4$, $\mu = m = 2$. 
The full AMR grid extends 
from $-100$ to $100$, while the inset shows the range $-13$ to $13$.} 
\label{C=-2.Q=-0.5.-100to100.tau=0.005.2inone}
\end{figure}
The contour levels crossing the $x$ axis between $-100$ and zero are, respectively, 
$\{1.0015, 1.005, 1.005, 1.0015, 0.96\}$.  The contour levels crossing the $y$ axis between 
$-100$ and zero are, respectively, $\{1.0015, 1.005, 1.01, 1.02, 1.1\}$.
The inset in Fig.~\ref{C=-2.Q=-0.5.-100to100.tau=0.005.2inone} shows the range in $x$ and $y$ 
from $-13$ to $13$.  In the inset figure, the contours crossing the $x$ axis between $-13$ to zero are 
$\{1.005, 1.0015, 0.96\}$ and the contours crossing the $y$ axis between $-13$ and zero are 
$\{1.01, 1.02, 1.1\}$. The function $u$ forms two peaks along the $y$ axis with maximum values 
$2.03$, and two valleys along the $x$ axis with minimum values $0.296$. The value of $u$ at the 
puncture is $1.10$. The profiles of $u$ 
along the three axes look similar to those shown in Fig.~\ref{RUN.AMR10.Mp=10.M0=2.Q=-.5.2D}.

With AMR, we are able to push the boundaries out quite far while maintaining
high resolution around the puncture.  For the data described above, the computational 
grid extends from $-100$ to $100$ and has
refined itself to reach a maximum relative truncation error of $0.008$.  The lowest
resolution region is equivalent to a $64^3$ grid with $\Delta x = 3.125$.  The highest
resolution region is equivalent to a $262,144^3$ grid with $\Delta x = 0.000763$.
\begin{figure}
\includegraphics[scale=1]{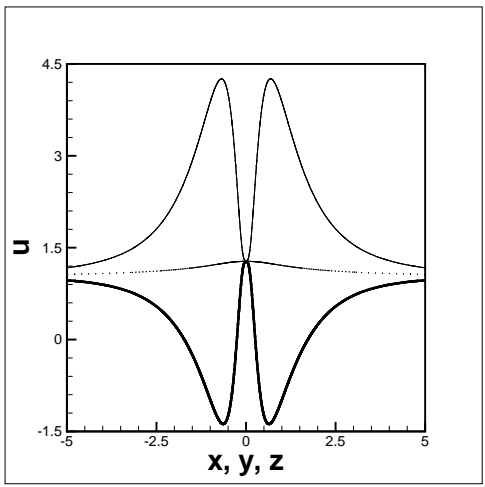}
\caption{The function $u$ plotted along the three coordinate axes for a non--isometric distorted black hole 
with  $c=-2$,  $Q_0 = -0.5$, $n = 4$, $\mu = 2$, $m = 10$. The thin solid line (top curve) is $u$ along the $y$--axis, 
the dotted line (middle curve) is $u$ along the
$z$--axis, and the thick solid line (bottom curve) is $u$ along the $x$--axis.}
\label{RUN.AMR10.Mp=10.M0=2.Q=-.5.2D}
\end{figure}

The puncture method allows us to define a new class of initial data: distorted black holes 
that are not isometric ($\mu \ne m$).  
Figure \ref{RUN.AMR10.Mp=10.M0=2.Q=-.5.2D} shows such a data set with $c=-2$, $Q_0 = -0.5$,  $n = 4$, $\mu = 2$, and 
$m = 10$. The results were computed on a grid extending from $-100$ to $100$ in each dimension. The grid 
spacing for the lowest resolution region, adjacent to the grid boundaries, is $\Delta x = 3.125$. The grid 
spacing for the highest resolution region, surrounding the puncture, is $\Delta x = 0.01221$. 
The ADM mass at infinity is $M_\infty = 10.7$, while the ADM mass at the puncture is $M_0 = 12.8$.  Figure
\ref{RUN.AMR10.Mp=10.M0=2.Q=-.5.2D} shows the behavior of the nonsingular part $u$ of the 
conformal factor along the coordinate axes.  The thin solid line plots $u$ along the $y$ axis, the thick solid line
plots $u$ along the $x$ axis, and the dotted line plots $u$ along the $z$ axis.  The two peaks in $u$ 
along the $y$--axis have maximum values $4.26$, and the two valleys along the $x$--axis have minimum values 
$-1.38$. The value of $u$ at the puncture is $1.28$. 

Note that for the non--isometric data described above, $u$ takes on negative values in two small regions 
of radius $\sim 1$ 
at locations $\sim \pm 1$ along the $x$ axis. However, for this data set, the conformal factor 
$\psi = u + m/(2r)$ is everywhere positive. We have explored other data sets that contain regions with $u<0$, and 
in each case we find that the conformal factor $\psi$ is positive everywhere. For data with $c=-2$,  $Q_0 = -0.5$, 
and $n = 4$, the most extreme data sets  we have studied  have ratios $\mu/m = 1/ 300$ 
and $\mu/m = 100$. For the case $\mu/m = 1/300$, the function $u$ reaches a minimum value of $\sim -120$ 
but the conformal factor remains positive. For the case $\mu/m = 100$ the function $u$, and 
therefore also $\psi$,  is always positive.


\begin{acknowledgments}
We would like to thank the numerical relativity group at NASA Goddard Space Flight Center, and especially Dae-Il Choi, 
for their help and support. We would also like to thank James Isenberg, Kevin Olson, and Peter MacNeice for 
helpful discussions.  This work was supported by NASA Space Sciences Grant ATP02-0043-0056. 
Computations were carried out 
on the North Carolina State University IBM Blade Center Linux Cluster.  
The PARAMESH software used in this work was developed at NASA Goddard Space Flight Center 
under the HPCC and ESTO/CT projects. 
\end{acknowledgments}
\bibliography{references}
\end{document}